\documentclass[aps,pra,twocolumn,superscriptaddress,10pt]{revtex4-2}
\usepackage{silence}
\usepackage{amsfonts}
\usepackage{eucal}
\usepackage{amsmath, amsthm, amssymb}
\usepackage{dsfont}

\DeclareFontFamily{OT1}{pzc}{}
\DeclareFontShape{OT1}{pzc}{m}{it}{<-> s * [1.150] pzcmi7t}{}
\DeclareMathAlphabet{\mathpzc}{OT1}{pzc}{m}{it}

\usepackage{hyperref}

\usepackage{float}

\begin{document}


\title{Clifford groups are not always 2-designs}


\author{Matthew A.\ Graydon}
\email[]{m3graydo@uwaterloo.ca}
\affiliation{Institute for Quantum Computing, University of Waterloo, Waterloo, Ontario N2L 3G1, Canada}
\affiliation{Department of Applied Mathematics, University of Waterloo, Waterloo, Ontario N2L 3G1, Canada}

\author{Joshua Skanes-Norman}
\affiliation{Institute for Quantum Computing, University of Waterloo, Waterloo, Ontario N2L 3G1, Canada}
\affiliation{Department of Applied Mathematics, University of Waterloo, Waterloo, Ontario N2L 3G1, Canada}
\affiliation{Keysight Technologies Canada, Kanata, ON K2K 2W5, Canada}

\author{Joel J.\ Wallman}
\affiliation{Institute for Quantum Computing, University of Waterloo, Waterloo, Ontario N2L 3G1, Canada}
\affiliation{Department of Applied Mathematics, University of Waterloo, Waterloo, Ontario N2L 3G1, Canada}
\affiliation{Keysight Technologies Canada, Kanata, ON K2K 2W5, Canada}

\date{\today}

\begin{abstract}
The Clifford group is the quotient of the normalizer of the Weyl-Heisenberg group in dimension $d$ by its centre. We prove that when $d$ is not prime the Clifford group is not a group unitary $2$-design. Furthermore, we prove that the multipartite Clifford group is not a group unitary 2-design except for the known cases wherein the local Hilbert space dimensions are a constant prime number. We also clarify the structure of projective group unitary $2$-designs. We show that the adjoint action induced by a group unitary $2$-design decomposes into exactly two irreducible components; moreover, a group is a unitary 2-design if and only if the character of its so-called $U\overline{U}$ representation is $\sqrt{2}$.
\end{abstract}

\maketitle

\section{Introduction}
The Clifford group features prominently in quantum information science. In their revolutionary work on quantum error correction, Calderbank, Rains, Shor, and Sloane \cite{Calderbank1998} adopted the term `Clifford group' citing \cite{Bolt1961I,Bolt1961II}. One can trace the etymology slightly further back to \cite{Horadam1957} and, with the right kind of eyes, see Clifford's algebras \cite{Clifford1871}. The Clifford group is the subgroup of the projective unitary group in dimension $d\in\mathbb{N}$ whose induced adjoint action preserves the Weyl-Heisenberg group \cite{Appleby2005}. The multipartite Clifford group is the quotient of the normalizer of tensor products of local Weyl-Heisenberg groups by its centre. 

Unitary 2-designs are structures at the heart of core quantum information protocols including randomized benchmarking \cite{Emerson2005,Helsen2020}, process tomography \cite{Chuang1997,Scott2008}, and cryptography \cite{Bennett1984,Ambainis2009}. We shall meet them formally in Section \hyperref[secDes]{II.B}. There exist many equivalent definitions, each with nice properties. One is that a unitary 2-design is a measurable set of unitary operators $U_{g}$ such that $\int_{\mathsf{G}}\int_{\mathsf{G}}|\mathrm{Tr}U_{g}^{\dagger}U_{h}|^{4}d\mu(g)d\mu(h)=2$ where the index set is, at this stage, only provocatively denoted by $\mathsf{G}$ and its measure by $\mu$. There are indeed fundamental connections between representation theory and unitary 2-designs. We shall equip ourselves for that discussion in Section \hyperref[secRep]{II.A}

Our final preliminary Section \hyperref[secWH]{II.C} presents the explicit definitions of Clifford, Pauli, and Weyl-Heisenberg groups. The Clifford group is a unitary $2$-design for prime dimensions \cite{DiVincenzo2002,Dur2005,Dankert2005,Gross2007,Dankert2009}. If each local Hilbert space dimension is a constant prime, then the multipartite Clifford group is a unitary 2-design and furthermore a unitary $3$-design when the aforementioned prime is two \cite{Zhu2017} (\textit{c.f.} \cite{Webb2016}.) The twirl of a quantum channel with respect to a unitary 2-design produces the unique depolarizing channel exhibiting the same average fidelity \cite{Nielsen1996}. Thus the multipartite Clifford group on, say, $n\in\mathbb{N}$ qubits can be used to design desirable channels. We then ask: are Clifford groups unitary 2-designs in all finite dimensions?

Theorem \hyperref[thm1]{III.1} and Theorem \hyperref[cor1]{III.2} establish negative answers. The Clifford group is not a unitary $2$-design when the dimension $d$ is a composite number, and the multipartite Clifford group is not a unitary $2$-design unless each local system has the same prime Hilbert space dimension. We would like to acknowledge that our results reach to the realm of folklore in certain (but not all \cite{Jafarzadeh2020,Goswami2021}) circles \cite{Gross2020}. In those circles, it will be enough for us to remark that Weyl-Heisenberg displacement operators exist with distinct orders, and therefore the tensor square of the Clifford group admits greater than two irreducible components. Our Section \hyperref[secDes]{III} formalizes the foregoing intuition given our observations in Section \hyperref[secDes]{II}. 

The Galoisian variant of the Clifford group defined by Appleby \cite{Appleby2009} and the related group defined by Chau \cite{Chau2005} are phrased in terms of linear operators on vector spaces over finite fields of order a prime power. These variants exist only in prime power dimensions, and they survive our analysis insofar as they are, although perhaps not naturally, isomorphic to multipartite Clifford groups on systems of a constant prime dimension. The Galoisian Clifford groups therefore happily evade our theorems.

Our proofs involve projective representation theory of compact groups \cite{Kirillov1976}. A projective group unitary 2-design is a unitary 2-design arising via the image of a suitable compact group under a projective unitary representation. Zhu \textit{et al}. \cite{Zhu2016} have developed a rich theory of projective group unitary 2-designs. Herein our Lemma \hyperref[lem2]{2} establishes further connections along these lines. We characterize projective group unitary 2-designs via conditions on their induced adjoint and so-called $U\overline{U}$ representations, which are indeed natural in the context of designing channels.

We remark that both the single system and multipartite Weyl-Heisenberg groups are group unitary $1$-designs in all finite dimensions. Indeed, they both act irreducibly on their underlying complex Hilbert spaces and so are unitary $1$-designs \cite{Roy2009}. Their normalizers contain them as normal subgroups, and so it follows that both the Clifford group and the multipartite Clifford group are also always group unitary $1$-designs. We draw further discussion and conclude in Section \hyperref[secDes]{IV}. 

\label{sec2}\section{Preliminaries}
\label{secRep}\subsection{Representations}
We shall now recall some elementary group theory \cite{Kirillov1976}. Let $\mathsf{G}$ be a compact (\textit{e.g}.\ a finite or matrix Lie) group. A subgroup $\mathsf{N}$ of $\mathsf{G}$ is normal when $\forall g\in\mathsf{G}$ it holds that $g\mathsf{N}g^{-1}=\mathsf{N}$, in which case one writes $\mathsf{N}\trianglelefteq\mathsf{G}$. The centre of $\mathsf{G}$ is $\mathrm{Z}_{\mathsf{G}}\equiv\{c\in\mathsf{G}\;|\;\forall g\in\mathsf{G}\;cg=gc\}\trianglelefteq\mathsf{G}$. The quotient group $\mathsf{G}/\mathrm{Z}_{\mathsf{G}}$ consists of all cosets $g\mathrm{Z}_{\mathrm{G}}\equiv\{gc\;|\;c\in\mathrm{Z}_{\mathsf{G}}\}$ with a group operation $g\mathrm{Z}_{\mathrm{G}}h\mathrm{Z}_{\mathrm{G}}\equiv gh\mathrm{Z}_{\mathrm{G}}$ for all $g,h\in\mathsf{G}$. Let $\mathsf{U}(\mathpzc{H})$ be the group of unitary automorphisms on a separable (\textit{i.e}.\ countably infinite or finite dimensional) complex Hilbert space $\mathpzc{H}$. The projective unitary group on $\mathpzc{H}$ is $\mathsf{PU}(\mathpzc{H})\equiv\mathsf{U}(\mathpzc{H})/\mathsf{U}(1)$ with the isomorphism $\mathrm{Z}_{\mathsf{U}(\mathpzc{H})}\cong\mathsf{U}(1)$ implicit.

A unitary representation of a compact group $\mathsf{G}$ on $\mathpzc{H}$ is a group homomorphism $\pi:\mathsf{G}\longrightarrow\mathsf{U}(\mathpzc{H})::g\longmapsto \pi(g)$.
A $\mathsf{G}$-invariant subspace of $\mathpzc{H}$ (with respect to $\pi$) is a subspace $\mathpzc{W}\subseteq\mathpzc{H}$ such that $\forall g\in\mathsf{G}$ and $\forall |\psi\rangle\in\mathpzc{W}$ $\pi(g)|\psi\rangle\in\mathpzc{W}$. A subrepresentation of $\pi$ on $\mathpzc{W}$ is a group homomorphism $\varphi:\mathsf{G}\longrightarrow\mathsf{U}(\mathpzc{W})$ regarding the image $\pi(g)$ as its formal restriction inside $\mathsf{U}(\mathpzc{W})$ for all $g\in\mathsf{G}$. A unitary representation $\pi$ is irreducible when its only $\mathsf{G}$-invariant subspaces are $\emptyset$ and $\mathpzc{H}$. A unitary representation of a compact group $\mathsf{G}$ splits into a direct sum of finite-dimensional irreducible unitary subrepresentations. The latter are in fact determined up to isomorphism by their characters. The character $\chi_{\pi}$ of a unitary representation $\pi$ is
\begin{equation}
\chi_{\pi}:\mathsf{G}\longrightarrow\mathbb{C}::g\longmapsto\mathrm{Tr}_{\mathpzc{H}}\pi(g)\text{.}
\end{equation} 
The inner product between two characters $\chi_{\phi}$ and $\chi_{\pi}$ is
\begin{equation}
\langle \chi_{\phi},\chi_{\pi}\rangle\equiv\int_{\mathsf{G}}\overline{\chi_{\phi}(g)}\chi_{\pi}(g)d\mu(g)\text{,}
\label{inProd}
\end{equation} 
with $\mu$ the normalized Haar measure on $\mathsf{G}$ and where an overline denotes complex conjugation. The induced norm is $\|\chi_{\pi}\|\equiv\sqrt{\langle\chi_{\pi},\chi_{\pi}\rangle}$. 
The multiplicity $m_{\varphi}$ of an irreducible unitary subrepresentation $\varphi$ in $\pi$ is $m_{\varphi}\equiv\int_{\mathsf{G}}\overline{\chi_{\varphi}(g)}\chi_{\pi}(g)d\mu(g)=\langle\chi_{\varphi},\chi_{\pi}\rangle$.

A projective unitary representation of a compact group $\mathsf{G}$ on $\mathpzc{H}$ is a group homomorphism 
\begin{equation}
\tilde{\pi}:\mathsf{G}\longrightarrow\mathsf{PU}(\mathpzc{H})::g\longmapsto \pi(g)\mathsf{U}(1)\text{.}
\label{repDef}
\end{equation}
From any such $\tilde{\pi}$ as in Eq.~\eqref{repDef} one can construct a proper unitary representation via its conjugate tensor square \footnote{The conjugate tensor square is the tensor product of two proper unitary representations in those cases wherein the associated 2-cocycle is trivial.} 
\begin{equation}
\pi\otimes\overline{\pi}:\mathsf{G}\longrightarrow\mathsf{U}(\mathpzc{H}\otimes\mathpzc{H})::g\longmapsto \pi(g)\otimes\overline{\pi(g)}\text{.}
\end{equation}
Indeed $\pi\otimes\overline{\pi}$ (the so-called $U\overline{U}$ representation) is proper since every projective unitary representation admits a 2-cocycle $\alpha:\mathsf{G}\times\mathsf{G}\longrightarrow\mathsf{U}(1)$ such that $\forall g,h\in\mathsf{G}$ $\pi(g)\pi(h)=\alpha(g,h)\pi(gh)$. In Appendix \hyperref[app]{A} we show that the conjugate tensor square is equivalent to the induced adjoint action 
\begin{eqnarray}
\text{Ad}_{\pi}:\mathsf{G}\longrightarrow\mathsf{U}\big(\mathfrak{B}(\mathpzc{H})\big)::g\longmapsto \text{Ad}_{\pi(g)}\text{,} 
\end{eqnarray}
where for all bounded linear operators $X\in\mathfrak{B}(\mathpzc{H})$ and $\forall g\in\mathsf{G}$ the action is $\text{Ad}_{\pi(g)}(X)\equiv \pi(g)X\pi(g)^{\dagger}$.
\label{secDes}\subsection{Designs}
Unitary $t$-designs are highly symmetric nets of unitary operators, in some sense covering the unitary group to mimic its Haar averages. A unitary $t$-design with $t\in\mathbb{N}$ in dimension $d\in\mathbb{N}$ is an indexed family $\{\pi(g)\}_{g\in\mathsf{G}}\subseteq\mathsf{U}(\mathpzc{H}_{d})$ endowed with a normalized measure $\mu$ such that for all $d^{t}\times d^{t}$ matrices $x$ with entries from $\mathbb{C}$ and with $\mu_{\text{H}}$ the unique normalized Haar measure on $\mathsf{U}(\mathpzc{H}_{d})$
\begin{equation}
\int_{\mathsf{G}}\pi(g)^{\otimes^{t}}\!x\big(\pi(g)^{\!\dagger}\big)^{\otimes^{t}}d\mu(g)=\int_{\mathsf{U}(\mathpzc{H}_{d})} U^{\otimes^{t}}\!x(U^{\!\dagger})^{\otimes^{t}} d\mu_{\text{H}}(U)\text{.}
\label{tdesdef}
\end{equation}

Unitary $t$-designs admit several equivalent definitions \cite{Dankert2009,Roy2009,Bengtsson2017}. We shall be interested in the case $t=2$. Strictly speaking, the following result in \cite{Gross2007} is stated within the context of finite sets; however and happily, their proof runs equally well in the continuous case.\\[0.2cm]
\label{lem1}\noindent\textbf{Lemma 1}[\cite{Gross2007}] \textit{Let} $\mu$ \textit{be a normalized measure on} $\pi(\mathsf{G})\equiv\{\pi(g)\}_{g\in\mathsf{G}}\subseteq\mathsf{U}(\mathpzc{H}_{d})$\textit{. Then} $\pi(\mathsf{G})$ \textit{is a unitary 2-design if and only if} $\int_{\mathsf{G}}\int_{\mathsf{G}}|\mathrm{Tr}\pi(g)^{\dagger}\pi(h)|^{4}d\mu(g)d\mu(h)=2$\textit{.}\\[0.2cm]
\indent A group unitary $2$-design is a subgroup of the unitary group $\mathsf{U}(\mathpzc{H}_{d})$ satisfying Eq.~\eqref{tdesdef} for $t=2$ \cite{Gross2007}. A rich theory has in fact been developed for all $t\in\mathbb{N}$ in the context of projective representations \cite{Zhu2017}. Let $\mathsf{G}$ be a compact group with normalized Haar measure $\mu$. In Appendix \hyperref[appB]{B} we, without difficulty, extend a technique from \cite{Gross2007} to prove that 
\begin{equation}
\|\chi_{\pi\otimes\overline{\pi}}\|^{2}=\int_{\mathsf{G}}\int_{\mathsf{G}}|\mathrm{Tr}\pi(g)^{\dagger}\pi(h)|^{4}d\mu(g)d\mu(h)\text{.}
\label{proveB}
\end{equation}

A projective group unitary 2-design is by definition a projective unitary representation $\tilde{\pi}$ as in Eq.~\eqref{repDef} such that the multiplicative extension 
\begin{equation}
\tilde{\pi}(\mathsf{G})\mathsf{U}(1)\equiv\{\pi(g)\gamma\;|\;g\in\mathsf{G}\wedge\gamma\in\mathsf{U}(1)\}\text{,} 
\end{equation}
is a group unitary 2-design. We now have the following. \\[0.2cm]
\label{lem2}\noindent\textbf{Lemma 2} \textit{Let} $\tilde{\pi}:\mathsf{G}\longrightarrow\mathsf{PU}(\mathpzc{H}_{d})$ \textit{be a projective unitary representation of a compact group} $\mathsf{G}$\textit{. Then the following statements are equivalent.}
\begin{enumerate}
\item $\tilde{\pi}(\mathsf{G})\mathsf{U}(1)$ \textit{is a group unitary 2-design.}
\item $\pi\otimes\overline{\pi}$ \textit{admits exactly two irreducible components.}
\item $\|\chi_{\pi\otimes\overline{\pi}}\|=\sqrt{2}\textit{.}$
\end{enumerate}
\noindent\textit{Proof.} The equivalence of $(1.)$ and $(3.)$ follows directly from Eq.~\eqref{proveB} (see Appendix \hyperref[appB]{B}) in the light of Lemma \hyperref[lem1]{1}. In order to see that $(2.)$ and $(3.)$ are equivalent, it suffices to remark that the norm of a character is the square root of the sum of the squares of the multiplicities of the irreducible components of the representation in question; crucially, within the given premises, that number is $\sqrt{2}$.
\begin{flushright}
\qed
\end{flushright}
\newpage
\label{secWH}\subsection{Groups}
The Pauli qubit operators in matrix form are
\begin{equation}
X\equiv\begin{pmatrix}0&\hspace{0.2cm}1\\ 1&\hspace{0.2cm}0\end{pmatrix},\hspace{0.2cm}Y\equiv\begin{pmatrix}0&-i\\ i&\hspace{0.2cm}0\end{pmatrix},\hspace{0.2cm}Z\equiv\begin{pmatrix}1&\hspace{0.2cm}0\\ 0&-1\end{pmatrix}\text{.}
\end{equation}
The Pauli group modulo its centre (namely, $\{\pm i\mathds{1}_{2},\pm \mathds{1}_{2}\}$) is $\{\mathds{1}_{2},X,Y,Z\}$ with $\mathds{1}_{2}$ the usual $2\times 2$ identity matrix. We denote the Pauli group by $\mathsf{P}$. This can be generalized to arbitrary finite dimensions as follows. 

The Weyl-Heisenberg group (in a finite dimension $d$) is denoted by $\mathsf{WH}_{d}$ and defined as follows. Let $\mathpzc{H}_{d}$ be a finite $d$-dimensional complex Hilbert space. Let $\{|e_{0}\rangle,\dots,|e_{d-1}\rangle\}$ be an orthonormal basis for $\mathpzc{H}_{d}$ and define the phase and shift operators, respectively, via their action on the foregoing basis as follows: for all $r\in\{0,1,\dots,d-1\}$ and with $\omega\equiv e^{2\pi i/d}$ the actions
\begin{eqnarray}
T|e_{r}\rangle&=&\omega^{r}|e_{r}\rangle\text{,}\\
S|e_{r}\rangle&=&|e_{r+1\text{ mod }d}\rangle\text{.}
\end{eqnarray}
Let $\tau\equiv-e^{i\pi/d}$, so that $\tau^{d^{2}}=1$ and $\omega=\tau^{2}$. For each $\vec{p}=(p_{1},p_{2})\in\mathbb{Z}_{d}^{2}$ we define, following Appleby \cite{Appleby2005}, the Weyl-Heisenberg displacement operators
\begin{equation}
D_{\vec{p}}=\tau^{p_{1}p_{2}}S^{p_{1}}T^{p_{2}}\text{.}
\end{equation}
Define $f(d)\equiv d$ if $d$ is odd and $f(d)\equiv 2d$ if $d$ is even. Next, introduce the following symplectic $\mathbb{Z}_{d}$-bilinear form
\begin{eqnarray}
\Omega&:&\mathbb{Z}_{d}^{2}\times\mathbb{Z}_{d}^{2}\longrightarrow\mathbb{Z}_{f(d)}\nonumber\\
&::&(\vec{p},\vec{q})\longmapsto\Omega(\vec{p},\vec{q})\equiv(p_{2}q_{1}-p_{1}q_{2})\text{ mod }f(d)\text{.}
\end{eqnarray} 
One can verify that  
\begin{eqnarray}
D_{\vec{p}}D_{\vec{q}}=\tau^{\Omega(\vec{p},\vec{q})}D_{\vec{p}+\vec{q}}\text{,}\label{projRep}\\
D_{\vec{p}}^{\dagger}=D_{-\vec{p}}\text{.} 
\end{eqnarray}
The Weyl-Heisenberg group $\mathsf{WH}_{d}$ (in dimension $d$) is herein defined to be generated by the Weyl-Heisenberg displacement operators. Therefore the cardinality is $|\mathsf{WH}_{d}|=d^{2}f(d)$ in all finite dimensions $d$ and explicitly we have
\begin{eqnarray}
\mathsf{WH}_{d}=\left\{\tau^{k}D_{\vec{p}}\;\big|\; k\in\mathbb{Z}_{f(d)}\;\text{and}\;\vec{p}\in\mathbb{Z}_{d}^{2}\right\}\text{.}
\end{eqnarray}
The Clifford group $\mathsf{C}_{d}$ is the normalizer of $\mathsf{WH}_{d}$ within the unitary group $\mathsf{U}(\mathpzc{H}_{d})$, modulo its centre. 
\begin{equation}
\mathsf{C}_{d}\equiv\{ C\in\mathsf{U}(\mathpzc{H}_{d})\;|\; C\mathsf{WH}_{d}C^{\dagger}=\mathsf{WH}_{d}\}/\mathsf{U}(1)\text{.}
\end{equation}
There is a generalization of the Weyl-Heisenberg group into the context of multiple quantum systems. Let $\mathbf{d}=\{d_{1},\dots,d_{n}\}$ be a multiset of $n\in\mathbb{N}$ naturals. We define the multipartite Weyl-Heisenberg group $\mathsf{WH}_{\mathbf{d}}$ (in dimension $d\equiv d_{1}\cdots d_{n}$ and with respect to $\mathbf{d}$) via
\begin{equation}
\mathsf{WH}_{\mathbf{d}}\equiv\left\{\bigotimes_{j=1}^{n}P_{j}\;\Bigg|\;\forall j\in\{1,\dots,n\}\; P_{j}\in\mathsf{WH}_{d_{j}}\right\}\text{.}
\end{equation}
The multipartite Clifford group $\mathsf{C}_{\mathbf{d}}$ (in dimension $d\equiv d_{1}\cdots d_{n}$ and with respect to $\mathbf{d}$) is the normalizer of $\mathsf{WH}_{\mathbf{d}}$ within $\mathsf{U}(\mathpzc{H}_{d_{1}\cdots d_{n}})$, modulo its centre.
\label{sec3}\section{Proofs}
Recall that the order (if it exists) of a unitary matrix $U\in\mathsf{U}(\mathpzc{H}_{d})$ is the smallest $r\in\mathbb{N}$ such that $U^{r}=\mathds{1}_{d}$.\\[0.15cm]

\label{lem3}\noindent\textbf{Lemma 3} \textit{Let} $d\in\mathbb{N}$ \textit{be a composite number. Then there exist at least three Weyl-Heisenberg displacement operators in} $\mathsf{WH}_{d}$ \textit{admitting distinct orders.}

\noindent\textit{Proof.} In all dimensions, the order of $\mathds{1}_{d}$ is unity. Now, our premise is that $d$ is not prime, and so we may choose $\vec{p},\vec{q}\in\mathbb{Z}_{d}^{2}$ as follows. First, the fundamental theorem of arithmetic yields $d=p_{1}^{r_{1}}\cdots p_{n}^{r_{n}}$ where $n\in\mathbb{N}$ and each $p_{j}$ is a prime distinct from the rest, with the powers $r_{j}\in\mathbb{N}$. If $n=1$, then choose $a=p_{1}\in\mathbb{Z}_{d}$ and $b\in\mathbb{Z}_{d}$ such that $b$ does not divide $p_{1}$. Then $D_{(a,0)}$ has order $d/p_{1}$ while $D_{(b,0)}$ has order $d$. If $d$ is not a prime power, then choose $\vec{p}=(p_{1},0)$ and $\vec{q}=(p_{2},0)\in\mathbb{Z}_{d}^{2}$ and consider the displacement operators $D_{\vec{p}}$ and $D_{\vec{q}}$. Eq.\eqref{projRep} yields the orders of $D_{(p_{1},0)}$ and $D_{(p_{2},0)}$; $d/p_{1}$ and $d/p_{2}$, respectively.
\begin{flushright}
\qed
\end{flushright}

\label{thm1}\noindent\textbf{Theorem III.1} \textit{Let} $d\in\mathbb{N}$ \textit{be a composite number. Then the Clifford group} $\mathsf{C}_{d}$ \textit{is not a unitary} $2$-\textit{design.}

\noindent\textit{Proof.} Fix a unitary representation $\phi:\mathsf{C}_{d}\longrightarrow\mathsf{U}\big(\mathfrak{B}(\mathpzc{H}_{d})\big)$ given by $\phi(C_{g})X\equiv C_{g}X C_{g}^{\dagger}$. The reader will recognize $\phi$ as the adjoint action inherited from the general linear group, which is equivalent to the action of $C_{g}\otimes\overline{C_{g}}$ on $\mathpzc{H}_{d}\otimes\mathpzc{H}_{d}$ within the representation ring (see Appendix \hyperref[app]{A} for a proof.) The irreducible components of $\phi$ are thus in $1:1$ correspondence with the irreducible components of the representation taking $C_{g}$ to $C_{g}\otimes\overline{C_{g}}$. Now, since $d$ is composite, Lemma \hyperref[lem3]{3} implies that there exist at least three distinct orders amongst the displacement operators in $\mathsf{WH}_{d}$. Recall the Clifford group, by definition, is such that $\forall C_{g}\in\mathsf{C}_{d}$ one has $C_{g}\mathsf{WH}_{d}C_{g}^{\dagger}=\mathsf{WH}_{d}$; however, if $D_{\vec{p}}$ has order $r$, then so does $UD_{\vec{p}}U^{\dagger}$, in fact, $\forall U\in\mathsf{U}(\mathpzc{H}_{d})$. In light of the linear independence of the displacement operators, there must therefore be at least 3 irreducible components of $\phi$. It remains only to recall Lemma \hyperref[lem2]{2}.
\begin{flushright}
\qed
\end{flushright}

\label{cor1}\noindent\textbf{Theorem III.2} \textit{Let} $\mathbf{d}\equiv\{d_{1},\dots,d_{n}\}$ \textit{be a multiset of} $\mathbb{N}$\textit{. Then the multipartite Clifford group} $\mathsf{C}_{\mathbf{d}}$ \textit{is not a unitary 2-design if there exist some} $d_{j}\neq d_{k}$ \textit{in} $\mathbf{d}$\textit{, nor is} $\mathsf{C}_{\mathbf{d}}$ \textit{a unitary 2-design if each} $d_{j}$ \textit{is the same composite number.}

\noindent\textit{Proof.} Immediate from the proof of Theorem \hyperref[thm1]{III.1} and the fact that $\text{span}_{\mathbb{C}}\mathsf{WH}_{\mathbf{d}}=\mathfrak{B}(\mathpzc{H}_{d})$ with $d=d_{1}\cdots d_{n}$.
\begin{flushright}
\qed
\end{flushright}
\label{sec4}\section{Discussion}
We have shown that the Clifford group is not a unitary $2$-design in composite dimensions (Theorem \hyperref[thm1]{III.1}); neither is the multipartite Clifford group unless each local dimension is a constant prime (Theorem \hyperref[cor1]{III.2}.)
\newpage
A natural question arising from our analysis is whether or not the tensor square $\pi\otimes\pi$ of a representation $\pi$ and its conjugate tensor square $\pi\otimes\overline{\pi}$ always admit irreducible components in $1:1$ correspondence. The answer is negative. Let $\pi:\mathbb{Z}_{4}\longrightarrow\mathsf{U}(2)::g\longmapsto\text{diag}(1,e^{i\pi g/2})$. Both the trivial and signed representations are enjoyed by $\pi\otimes\pi$ and $\pi\otimes\overline{\pi}$; however, the former admits two copies of $\varphi:\mathbb{Z}_{4}\longrightarrow\mathbb{C}::g\longmapsto\text{exp}(i\pi g/4)$, while the latter admits just one copy together with the complex conjugate of $\varphi$.

Finally, we remark that unitary $2$-designs indeed exist in all finite dimensions \cite{Seymour1984}. The Clifford group, however, is not always one of them. 

\appendix
\label{app}\section{The Conjugate Tensor Square and Adjoint Representations are Equivalent}
In this appendix we show that the conjugate tensor square and adjoint representations are equivalent within the representation ring. We now remind the reader that representations of compact groups are determined up to isomorphisms by their characters. Automatically, then, the multiplicities of irreducible components of equivalent representations coincide. We shall need the following.\\[0.3cm]
\label{propA1}\textbf{\textit{Proposition A.1}} \textit{Let} $\mathpzc{H}$ \textit{be a finite dimensional complex Hilbert space. Let} $U$ \textit{be a unitary operator on} $\mathpzc{H}$\textit{. Then}
\begin{equation}
\mathrm{Tr}_{\mathpzc{H}\otimes\mathpzc{H}}U\otimes\overline{U}=\mathrm{Tr}_{\mathfrak{B}(\mathpzc{H})}\text{Ad}_{U}\textit{.}
\end{equation}
\textit{Proof.} Let $d=\text{dim}_{\mathbb{C}}\mathpzc{H}$. Now, choose an orthonormal basis $\{|e_{1}\rangle,\dots,|e_{d}\rangle\}$ for $\mathpzc{H}$. Next,  construct $\big\{|E_{r,s}\rangle\!\rangle\equiv|e_{r}\rangle\langle e_{s}|\;\big|\;r,s\in\{1,\dots,d\}\big\}$, an orthonormal basis for $\mathfrak{B}(\mathpzc{H})$. Recall the definition $\text{Ad}_{U}(X)\equiv UXU^{\dagger}$. Thus equipping $\mathfrak{B}(\mathpzc{H})$ with the Hilbert-Schmidt inner product we calculate
\begin{eqnarray}
& &\mathrm{Tr}_{\mathfrak{B}(\mathpzc{H})}\text{Ad}_{U}\\ 
&=&\label{traceDef}\sum_{r=1}^{d}\sum_{s=1}^{d}\langle\!\langle E_{r,s}|UE_{r,s}U^{\dagger}\rangle\!\rangle\\ 
&=&\label{HS}\sum_{r=1}^{d}\sum_{s=1}^{d}\mathrm{Tr}(E_{r,s}^{\dagger}UE_{r,s}U^{\dagger})\\ 
&=&\label{breakBig}\sum_{r=1}^{d}\sum_{s=1}^{d}\langle e_{r}|Ue_{r}\rangle\langle e_{s}|U^{\dagger}e_{s}\rangle\\[0.1cm] 
&=&\label{traceProp1}\mathrm{Tr}_{\mathpzc{H}}U\mathrm{Tr}_{\mathpzc{H}}U^{\dagger}\\[0.3cm] 
&=&\label{traceProp2}\mathrm{Tr}_{\mathpzc{H}\otimes\mathpzc{H}}U\otimes\overline{U}\text{.} 
\end{eqnarray}
Eq.~\eqref{traceDef} and Eq.~\eqref{HS} are consequences of the definitions of the trace, the adjoint action $\text{Ad}_{U}$, and the Hilbert-Schmidt inner product. Eq.~\eqref{breakBig} is a direct calculation. Finally, Eq.~\eqref{traceProp1} and Eq.~\eqref{traceProp2} are simple consequences of elementary linear algebra; in particular, the trace under transposition.
\begin{flushright}
$\qed$
\end{flushright}
\textbf{\textit{Corollary A.2}} \textit{Let} $\tilde{\pi}:\mathsf{G}\longrightarrow\mathsf{PU}(\mathpzc{H})::g\longmapsto\pi(g)\mathsf{U}(1)$ \textit{be a projective unitary representation of a compact group} $\mathsf{G}$\textit{. Then, within the representation ring,}  $\pi\otimes\overline{\pi}\cong\text{Ad}_{\pi}$\textit{.}\\[0.3cm]
\textit{Proof.} The character 
\begin{equation}
\chi_{\pi\otimes\overline{\pi}}:\mathsf{G}\longrightarrow\mathbb{C}::g\longmapsto\mathrm{Tr}_{\mathpzc{H}\otimes\mathpzc{H}}\pi(g)\otimes\overline{\pi(g)}
\end{equation} 
of $\pi\otimes\overline{\pi}$ is the sum of its irreducible characters weighted by their multiplicities, and all such irreducible characters trace over finite dimensional complex Hilbert spaces. It remains to recall Proposition \hyperref[propA1]{A.1}.
\begin{flushright}
$\qed$
\end{flushright}
\label{appB}\section{Proof of Eq.(7)}
Gross, Audenaert, and Eisert employ a technique in \cite{Gross2007} to prove that given a proper unitary representation $\pi$ of a finite group $\mathsf{G}$, $\pi(\mathsf{G}$) is a unitary $2$-design if and only if $\|\chi_{\pi\otimes\pi}\|=\sqrt{2}$. Their proof can without difficulty be generalized to the case of compact groups. Here we provide a record of how the aforementioned technique can be easily applied to prove the following.\\[0.2cm]

\noindent\textbf{\textit{Proposition B.1}} \textit{Let} $\tilde{\pi}:\mathsf{G}\longrightarrow\mathsf{PU}(\mathpzc{H})$ \textit{be a projective unitary representation of a compact group} $\mathsf{G}$\textit{. Then}
\begin{equation}
\|\chi_{\pi\otimes\overline{\pi}}\|^{2}=\int_{\mathsf{G}}\int_{\mathsf{G}}|\mathrm{Tr}\pi(g)^{\dagger}\pi(h)|^{4}d\mu(g)d\mu(h)\text{.}
\end{equation}
\textit{Proof.} We calculate
\begin{eqnarray}
&&\int_{\mathsf{G}}\int_{\mathsf{G}}|\mathrm{Tr}\pi(g)^{\dagger}\pi(h)|^{4}d\mu(g)d\mu(h)\\
\label{homProp}&=&\int_{\mathsf{G}}\int_{\mathsf{G}}|\alpha(g^{-1},h)\mathrm{Tr}\pi(g^{-1}h)|^{4}d\mu(g)d\mu(h)\\
\label{cocyle}&=&\int_{\mathsf{G}}\int_{\mathsf{G}}|\mathrm{Tr}\pi(g^{-1}h)|^{4}d\mu(g)d\mu(h)\\
\label{HaarProp}&=&\int_{\mathsf{G}}\int_{\mathsf{G}}|\mathrm{Tr}\pi(g^{-1}h)|^{4}d\mu(g)d\mu(g^{-1}h)\\
\label{FubiniProp}&=&\int_{\mathsf{G}}d\mu(g)\int_{\mathsf{G}}|\mathrm{Tr}\pi(x)|^{4}d\mu(x)\\
\label{HaarNorm}&=&\int_{\mathsf{G}}|\mathrm{Tr}\pi(x)|^{4}d\mu(x)\\
\label{alg1}&=&\int_{\mathsf{G}}\mathrm{Tr}\overline{\big(\pi(x)\otimes \pi(x)\big)}\mathrm{Tr}\big(\pi(x)\otimes \pi(x)\big)d\mu(x)\\
\label{last}&=&\|\chi_{\pi\otimes\overline{\pi}}\|^{2}\text{.}
\end{eqnarray}
Eq.~\eqref{homProp} is the group homomorphism property of $\pi$; Eq.~\eqref{cocyle} follows since the 2-cocycle has codomain $\mathsf{U}(1)$; Eq.~\eqref{HaarProp} applies translational invariance of the Haar measure; Eq.~\eqref{FubiniProp} is permitted by Fubini's Theorem; Eq.~\eqref{HaarNorm} follows since the Haar measure is normalized; Eq.~\eqref{alg1} is a consequence of trace properties of tensor products; and finally Eq.~\eqref{last} is a simple consequence of the definition. 
\begin{flushright}
$\qed$
\end{flushright}
\newpage

\begin{acknowledgments}
MG would like to thank Marcus Appleby for discussing what problems herein were open.
This research was undertaken thanks in part to funding from the Canada First Research Excellence Fund, the Government of Ontario, and the Government of Canada through NSERC.
\end{acknowledgments}

\bibliography{GSNW}

\end{document}